\documentclass[%
 reprint,
superscriptaddress,
amsmath,amssymb,
aps,
pra,
]{revtex4-1}
\usepackage{caption}
\usepackage{subcaption}
\usepackage{graphicx}
\usepackage{xcolor}
\usepackage{graphicx}
\usepackage{dcolumn}
\usepackage{bm}

\begin{document}

\title{Rotationally displaced electric field intensity distribution  around square nanoantennas induced by circularly polarized light}

\preprint{APS/123-QED}

\author{Naoki Ichiji*}
\affiliation{Institute of Industrial Science, The University of Tokyo, 4-6-1 Komaba, Meguro-Ku, Tokyo 153-8505, Japan}
\email{ichiji@iis.u-tokyo.ac.jp}
\author{Takuya Ishida}
\affiliation{Institute of Industrial Science, The University of Tokyo, 4-6-1 Komaba, Meguro-Ku, Tokyo 153-8505, Japan}
\author{Ikki Morichika}
\affiliation{Institute of Industrial Science, The University of Tokyo, 4-6-1 Komaba, Meguro-Ku, Tokyo 153-8505, Japan}
\author{Tetsu Tatsuma}
\affiliation{Institute of Industrial Science, The University of Tokyo, 4-6-1 Komaba, Meguro-Ku, Tokyo 153-8505, Japan}
\author{Satoshi Ashihara}
\affiliation{Institute of Industrial Science, The University of Tokyo, 4-6-1 Komaba, Meguro-Ku, Tokyo 153-8505, Japan}


\begin{abstract}
An optical field around regular polygon metal nanostructures excited by circularly polarized light can exhibit rotationally displaced intensity distributions. Although this phenomenon has been recognized, its underlying mechanisms has not been sufficiently  explained. Herein, finite-difference time-domain simulations and model analyses reveal that the rotationally displaced optical intensity  distribution can be generated when each of the linear polarization components that constitute circular polarization excites a superposition of multiple modes. The proposed model reasonably explains the rotationally displaced patterns for a square nanoantenna and other regular-polygon nanoantennas.
\end{abstract}

\maketitle

\section{Introduction}
Localized collective oscillations of surface electrons on metal nanostructures are acutely sensitive to the size and shape of the nanostructure and exhibit strong electric-field-enhancements~\cite{Ghenuche08PRL,Hanke12NL}.
Well-designed plasmonic antennas have been used to modulate various optical properties~\cite{Giannini11ChemRev}. 
In particular, the interaction between circularly polarized (CP) light and plasmonic structures has been extensively investigated~\cite{Lehmuskero13NL,Lee14NanoP,Cheng20NP,Hentschel17SciAd,Mun20LSA,Hu22LPR}.
During initial research, the enhancement of optical properties originated from the circular polarization, represented by optical chirality~\cite{Tang10PRL, Bliokh11PRA}, has primarily demonstrated by chiral structures lacking mirror-image congruency~\cite{Hentschel12NL,Duan16NL,Schaeferling12PRX}. 
However, recent studies have reported that optical chirality can be observed in the localized near-fields at the vertices of rectangle plasmonic structures excited by linearly polarized (LP) light~\cite{Schaeferling12OE,Hashiyada18ACSP,Okamoto19JMCC}. 

\begin{figure}[b!]
  \begin{center}
  \includegraphics[width=8.6cm]{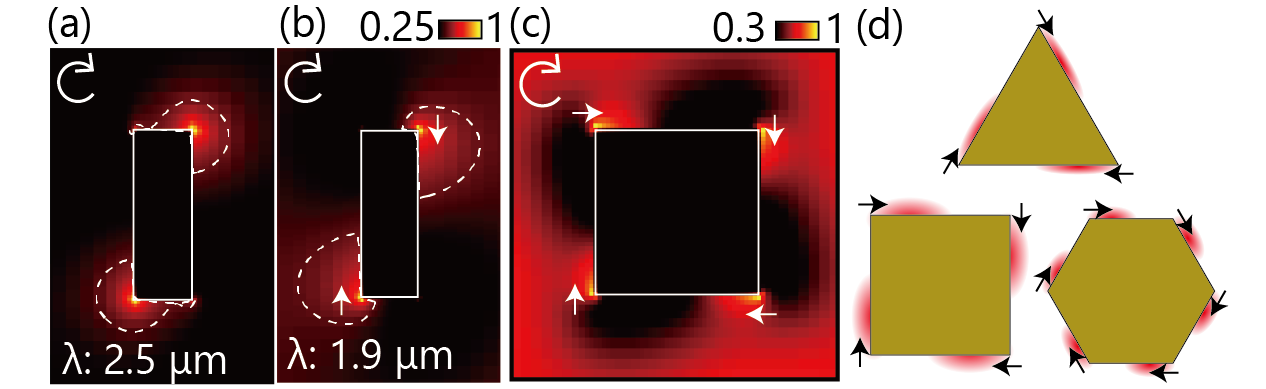}
  \end{center}
  \caption{Localization of the electric field on a nanoantenna under circularly polarized (CP) light: (a-c) Electric field intensity from electromagnetic simulations for (a,b) rectangular at different wavelength and (c) square nanoantenna. White dashed lines representing the intensity contour lines. (d) Schematic of the electric field distributions of regular polygonal nanoantennas.}
  \label{Fig:image}
\end{figure}

Furthermore, it has been shown that the spatial distribution of the electric field of an achiral structure excited by CP light is influenced by the handedness of the incident light~\cite{Hashiyada14JPCC, Horrer20NL, Okamoto19JMCC}, and has been applied to photochemical fabrication of chiral nanostructures~\cite{Saito18NL, Morisawa20ACSNano} and enantioselective sensing or trapping of chiral molecules~\cite{Liu22OE, Yamane23OE}. In rectangular nanoantennas, the handedness-dependent distribution appears as the localization of the electric field at one of the two pairs of diagonal vertices (Fig.~\ref{Fig:image} (a)). This selective localization has been explained by the superposition of the two fundamental plasmon modes associated with the major and minor axes~\cite{Oshikiri21ACSNano,Zu19NL}. However, it is also recognized that the time-averaged electric field intensity around a rectangular nanoantenna may exhibit a rotationally displaced spatial distribution depending on the exciting wavelength (Fig.~\ref{Fig:image} (b))~\cite{Saito18NL}; this wavelength-dependent two-dimensional spatial distribution cannot be explained solely by the fundamental mode.

A similar issue has been recognized for square nanoantennas, whereas the electric field intensity distribution around the square nanoantenna is isotropic at wavelengths longer than the dipole resonance wavelength, several studies reported that the intensity distribution becomes displaced in the direction of the electric field rotation of the incident CP light at shorter wavelengths (Fig.~\ref{Fig:image} (c)). This rotational displacement holds for other regular polygons~\cite{Homma23CNM, Shimomura20APL, Ishida23APL,Besteiro21NL} as shown in Fig.~\ref{Fig:image} (d), and structures with equivalent rotational symmetry~\cite{Horrer20NL}, and has been used to fabricate chiral structures~\cite{Homma23CNM, Shimomura20APL, Ishida23APL}.  
Nonetheless, the underlying mechanism of this rotational displacement remains unexplored.

In this study, we investigate the origin of the rotationally displaced electric field intensity distribution around a square nanoantenna excited by CP light, employing finite-difference time-domain (FDTD) simulations and model analyses. 
We find that rotationally displaced intensity distribution appears when the near-field excited by each of the linear polarization components that constitute circular polarization exhibits non-uniform phase distribution. The non-uniform phase distribution cannot arise from a single plasmon mode but arises from a superposition of multiple plasmon modes. 
Each of the two orthogonal linear polarization components excites a near field with the same spatial distribution but rotated 90 degrees from each other. When these two near fields are overlapped with a phase difference of $\pi/2$, the intensity distribution exhibits rotational displacement with respect to the square shape of the nanoantenna. We verify that the non-uniform phase distribution plays an important role for the emergence of the rotationally displaced intensity distribution around regular polygons.

\begin{figure}[t!]
  \begin{center}
  \includegraphics[width=8.6cm]{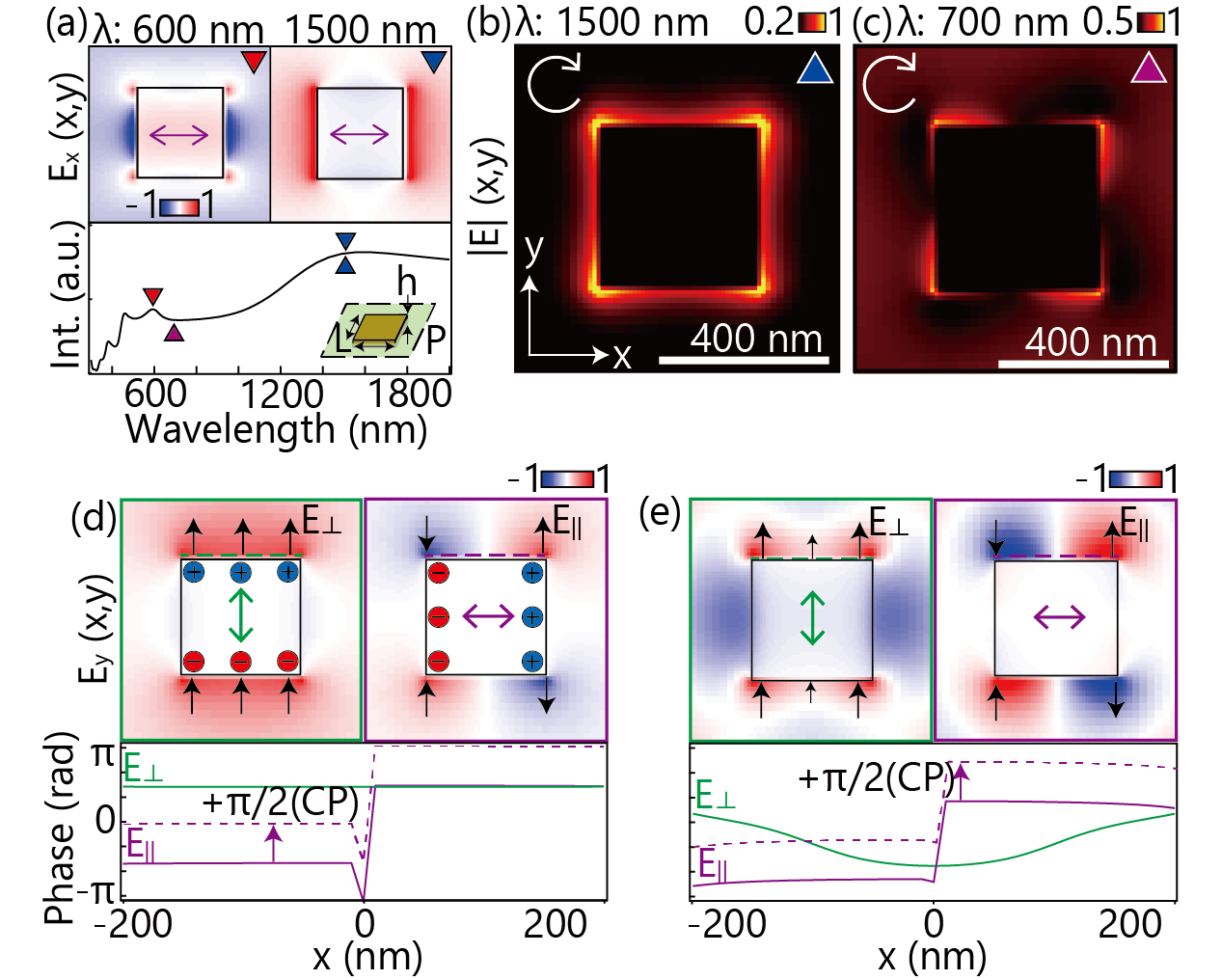}
  \end{center}
  \caption{(a) Scattering spectra of nanoantenna excited by $x$-polarized light. Top panels indicate the $E_{x}$ components at a representative phase. (b,c) Spatial distributions of the electric field intensity obtained at plane $P$. (d, e) $E_{y}$ components at a representative phase excited by the $y$-polarized (green frames) and $x$-polarized (purple frames) light. Phase profiles traced along the dashed lines are plotted in the bottom panels.}
  \label{Fig:FDTD}
\end{figure}
\section{FDTD simulation}
In our simulation, square gold (Au) nanoantenna in a vacuum with a thickness ($h$) of 30 nm and an edge length ($L$) of 400 nm is excited via normal incident light. The dielectric function of Au is obtained from the Palik database~\cite{Palik}. A minimum mesh width is 10 nm. Observation plane $P$ is set in the middle of the Au nanoantenna. All simulations were performed using ANSYS Lumerical (2023 R1).

Nanoantennas with a finite area are known to exhibit multiple resonant modes that depend on their size and shape~\cite{Zhang11NL,Pellarin16ACSN,Schmidt14NL}. The scattering spectrum in Fig.~\ref{Fig:FDTD} (a) confirms the existence of multiple resonance peaks. The top panels depict the spatial distributions of the $E_{x}$ components at a representative phase around the nanoantenna excited by $x$-polarized light. The resonance at 1500 nm indicates a clear dipole mode, whereas that at 600 nm displays a hexapole mode.

The intensity distribution of the electric field in the nanoantenna under CP light excitation exhibits a wavelength-dependent spatial pattern; Figs.~\ref{Fig:FDTD}(b) and \ref{Fig:FDTD}(c) illustrate the spatial distributions of the selected wavelengths. By contrast, the spatial distribution at the resonance wavelength of the dipole mode (1500 nm) is isotropic, and a rotationally displaced pattern is observed at 700 nm.
To investigate the factors responsible for the rotational displacement, we analyzed the near field under each of the $x$- and $y$-polarized light excitations, that constitute the CP light.

Fig.~\ref{Fig:FDTD} (d) depicts the electric field distribution $E_{y}(x,y)$ at a representative phase at $\lambda\!=\!1500$ nm. Hereafter, the electric fields excited by the light polarized perpendicular and parallel to an edge are referred to as $E_{\perp}$ and $E_{\parallel}$, respectively. 
The subscripts indicate the relationship between the edge and the polarization of the incident light. The directions of the calculated electric field indicated by black arrows (Figs.~\ref{Fig:FDTD} (d,e)) are consistent with those expected from the positions of the electrons for the dipole mode plotted schematically in the same figure. 
The phase distributions at the upper edge, indicated by the dashed horizontal lines, are plotted as solid lines in the bottom panel in Figs.~\ref{Fig:FDTD} (d,e). The phase distributions of $E_{\perp}$ and $E_{\parallel}$ are consistent with an intuitive oscillation pattern inferred from the dipole mode: $E_{\perp}$ is constant across the edge, and $E_{\parallel}$ is flat except for a $\pi$ shift on the $x\!<\!0$ and $x\!>\!0$ region, reflecting the sign difference of the electric field. Therefore, the phase difference between $E_{\perp}$ and $E_{\parallel}$ exhibits only discrete values of 0 or $\pi$.

By contrast, a nontrivial phase distribution is observed at wavelengths that shows a rotationally displaced pattern under CP light irradiation. As indicated in Fig.~\ref{Fig:FDTD} (e), the phase distributions of $E_{\perp}$ and $E_{\parallel}$ excited at a wavelength of 700 nm, particularly $E_{\perp}$, exhibit non-uniform curvetures. 
Consequently, a phase difference between $E_{\perp}$ and $E_{\parallel}$ is a non-discrete value, which is neither 0 nor $\pi$. 
This phase difference breaks the symmetry of the interference on each edge under CP light incidence. Considering the phase difference of $\pi/2$ between the orthogonal linear polarization components in CP light, the phase distribution of $E_{\parallel}$ relative to $E_{\perp}$ for CP light is positioned as indicated by the dashed purple line in Fig.~\ref{Fig:FDTD} (e). Consequently, $E_{\perp}$ and $E_{\parallel}$ interfere constructively on the left-hand side and vice versa on the right-hand side, inducing asymmetric intensity distribution along the edges. 

\section{Model calculation}
\begin{figure}[t!]
  \begin{center}
  \includegraphics[width=8.6cm]{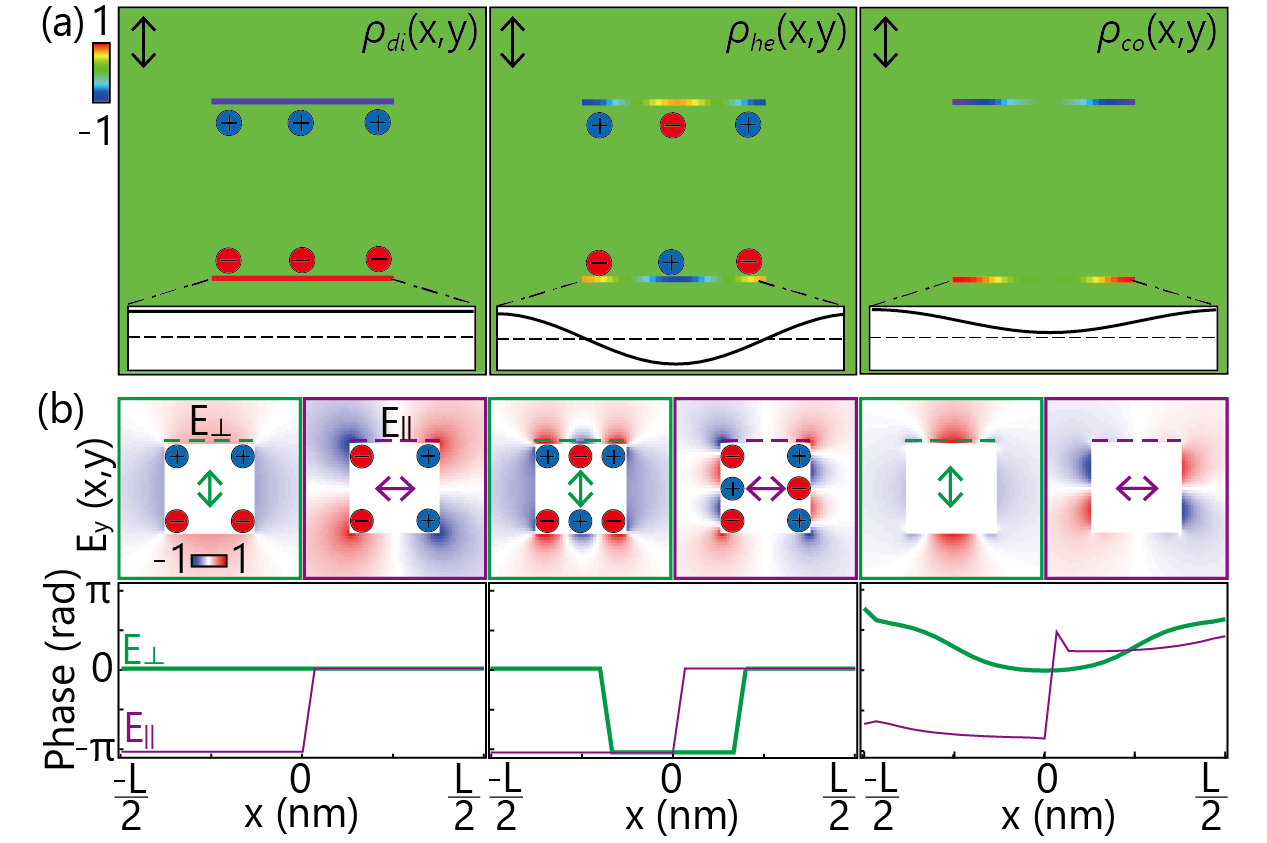}
  \end{center}
  \caption{(a) Charge density distributions for the dipole mode (left), hexapole mode (middle), and combined excitation (right) comprising both the dipole and hexapole modes. (b) Calculated $E_{y}$ components at a representative phase assuming $x$- and $y$- polarization of the excitation light. The bottom panels display the phase profiles traced along the dashed lines indicated in the top panel.
}
  \label{Fig:charge}
\end{figure}
As a single dipole mode is incapable of producing a distorted phase distribution, the non-uniform phase distribution can be attributed to a superposition of multiple resonant modes. To examine the validity of this hypothesis, we consider the superposition of charge distributions for each resonant mode. Here, the dipole mode and the hexapole mode, whose resonance peaks are observed on both sides of 700 nm as shown in Fig.~\ref{Fig:FDTD} (a), are used as the unit oscillation modes. The charge densities excited by the $y$-polarized light, which are $\rho_{\mathrm{di}}$ for the dipole mode, $\rho_{\mathrm{he}}$ for the hexapole mode, and $\rho_{\mathrm{co}}$ for the combined excitation, can be expressed in simplified forms as follows:
\begin{align}
&\rho_{\mathrm{di}}\left(y\!=\!\pm \frac{L}{2}; \phi(t)\right)\!=\!\pm \mathrm{cos}(\phi(t));\\
&\rho_{\mathrm{he}}\left(y\!=\!\pm \frac{L}{2}, x; \phi(t)\right)\!=\!\pm \mathrm{cos}\left(\frac{2\pi x}{L}\right)\mathrm{cos}(\phi(t));\\
&\rho_{\mathrm{co}}(\phi(t))\!=\!A_{\mathrm{di}}\rho_{\mathrm{di}}(\phi(t))+\!A_{\mathrm{he}}\rho_{\mathrm{he}}(\phi(t)+\phi_{\mathrm{diff}}),
\end{align}
where $\phi(t)$ denotes the temporal phase, and $A_{\mathrm{di}}$ and $A_{\mathrm{he}}$ indicate the scalar coefficients, with both set a value of 0.5. Additionally, $\phi_{\mathrm{diff}}$ represents the phase difference between the two modes. As the resonant wavelengths of distinct resonant modes are different, they have different phase delay with respect to the external field. Fig.~\ref{Fig:charge} (a) depicts the charge-density distribution for each mode. Assuming a phase delay of $\pi$ for the dipole mode and $\pi/2$ for the hexapole mode, $\phi_{\mathrm{diff}}$ is set to $\pi/2$. The corresponding charge distributions for the $x$-polarized light can be obtained by rotating each distribution by $90^\circ$.

Fig.~\ref{Fig:charge} (b) illustrates the electric field distribution obtained by applying Coulomb’s law to each charge density distribution. The phase distributions of $E_{\perp}$ and $E_{\parallel}$ obtained along the upper edge (dashed lines in the top panel) are presented in the bottom panels . The phase distributions of $E_{\perp}$ directly reflect their charge distributions across the bottom edge, whereas the phase distributions of $E_{\parallel}$ are dominated by the charge distribution at both sides of the edges. Therefore, $E_{\perp}$ of the dipole and hexapole modes is distinct, whereas $E_{\parallel}$ is similar. Consequently, $E_{\perp}$ in the combined excitation exhibits a phase distribution with a large concave at the center portion and a significantly larger degree of curvature in comparison with $E_{\parallel}$. The different curvatures of the phase distributions of $E_{\perp}$ and $E_{\parallel}$ on the same edge introduce a non-discrete phase difference between them that would never appear in the single mode.

\section{Discussion}
\begin{figure}[t]
  \begin{center}
  \includegraphics[width=8.6cm]{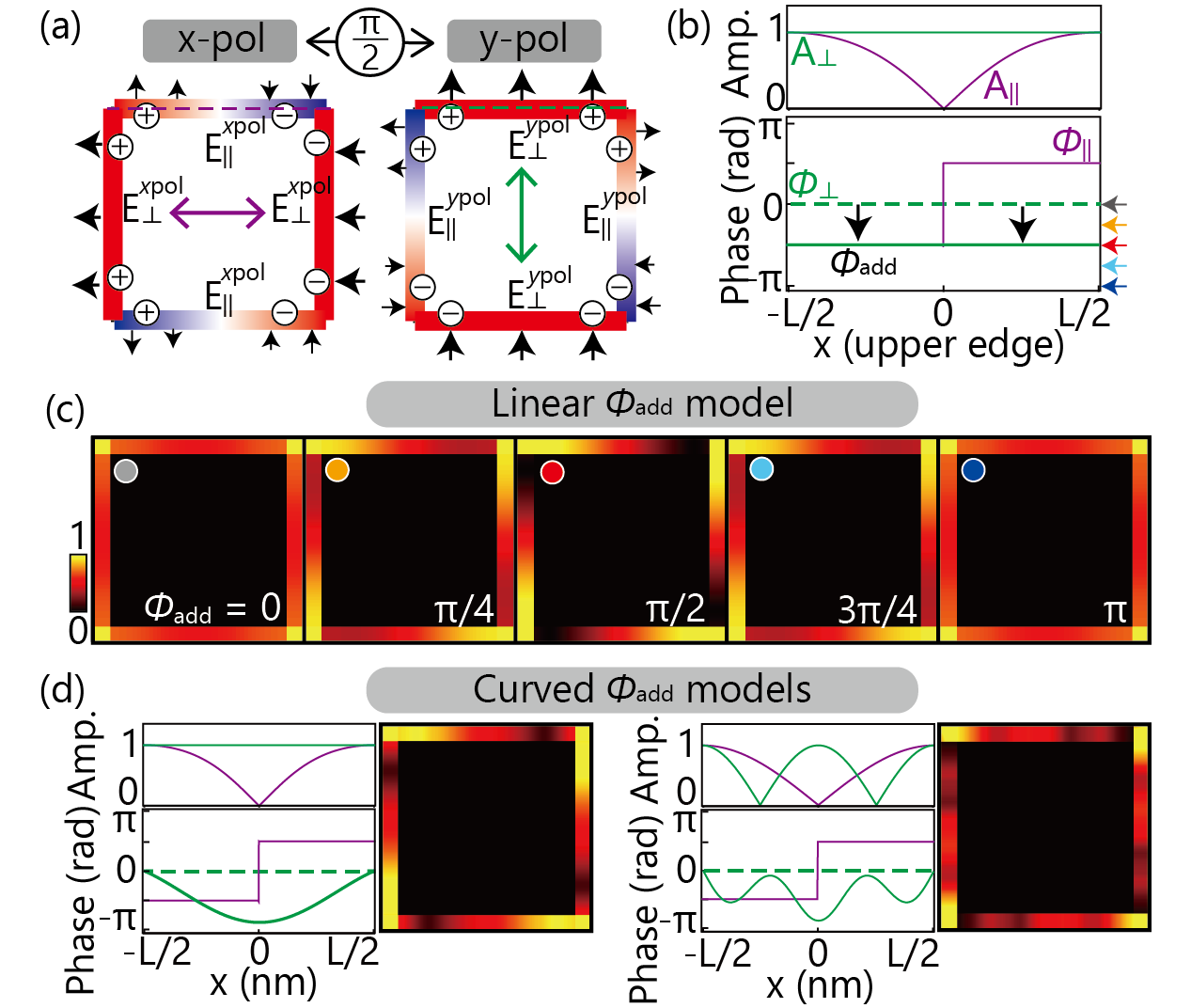}
  \end{center}
  \caption{ (a) Schematic of the electric field distribution in the model. Red and blue regions on the frames represent positive and negative electric fields, respectively. (b) Amplitude (top) and phase (bottom) distributions of the upper edge of the model under circularly polarized (CP) light excitation. (c) Calculated field distribution for typical $\phi_{\mathrm{add}}$. (d) Spatial electric field distributions calculated using the curved $\phi_{\mathrm{add}}$ models.}
  \label{Fig:model}
\end{figure}
The simulation results indicate that the non-uniform phase distributions along the edges are crucial for understanding the rotationally displaced pattern. However, it is not self-evident whether the fundamental origin of the rotationally displaced intensity distribution lies in the concave distribution of $E_{\perp}$ itself, or in the phase difference between $E_{\perp}$ and $E_{\parallel}$.
Here, we discuss the underlying mechanism of the rotationally displaced intensity distribution by employing a model with simplified spatial electric field distributions. The electric field components excited by the $x$- and $y$-polarized components, $E^{x\mathrm{pol}}$ and $E^{y\mathrm{pol}}$, respectively, are modeled assuming the electron distribution as depicted in Fig.~\ref{Fig:model} (a)~\cite{Matthaiakakis22AOM}. $E^{x\mathrm{pol}}$ at each side can be defined as follows:
\begin{align}
E_{\perp}^{x\mathrm{pol}} \left(x\!=\!\pm\frac{L}{2}, y;\phi(t)\right) &= A_{\perp}(y)\cos(\phi(t)+\phi_{\perp}(y));\\
E_{\parallel}^{x\mathrm{pol}} \left(x,y\!=\!\pm\frac{L}{2};\phi(t)\right) &= A_{\parallel}(x)\cos(\phi(t)+\phi_{\parallel}(x)),
\end{align}
where $A_{\perp}$ and $A_{\parallel}$ represent the amplitudes; $\phi_{\perp}$ and $\phi_{\parallel}$ indicate the phase distributions for $E_{\perp}$ and $E_{\parallel}$, respectively; and $\phi(t)$ denotes a temporal phase of the oscillation. We assumed a uniform electric field as $E_{\perp} (A_{\perp} =A_{1})$ and a sinusoidal electric field distribution as $E_{\parallel} (A_{\parallel}\!=\!|A_{2}\sin(2\pi x/L)|))$, where $A_{1}$ and $A_{2}$ are scalar coefficients set to 1. The phase distributions $\phi_{\perp}$ and $\phi_{\parallel}$ are simplified as
\begin{align}
\phi_{\perp}\left(x\!=\!\pm\frac{L}{2},y\right) &= -\phi_{\mathrm{add}};\\
\phi_{\parallel}\left(x,y\!=\!\pm \frac{L}{2}\right) &= \pm\pi\Theta(\pm x),
\end{align}
where $\Theta(x)$ denotes the step function and is 0 if $x\!<\!0$ and 1 if $x\!>\!0$, and $\phi_{\mathrm{add}}$ represents the hypothetical phase difference introduced between $E_{\perp}$ and $E_{\parallel}$. 
This coordinate-independent hypothetical phase difference is introduced for a generalized discussion that concentrates solely on the phase difference. $E^{y\mathrm{pol}}$ is defined in the same manner. Considering the $\pi/2$ phase difference between the orthogonal linear polarization components for the CP light, the field distribution excited by $E^{CP}$ can be expressed as
\begin{align}
E^{CP}(\phi(t),\phi_{\mathrm{add}}) &= E^{x\mathrm{pol}}\left(\phi(t)+\frac{\pi}{2}\right)+ E^{y\mathrm{pol}}(\phi(t)).\label{Eq:LC}
\end{align}

Fig.~\ref{Fig:model} (b) depicts the amplitude and phase distributions at the upper edge of the model ($y\!=\!L/2$) under CP light excitation. 
In the lower panel of Fig.~\ref{Fig:model} (b), assigning a positive constant to $\phi_{\mathrm{add}}$ corresponded to moving the green line downwards. This shift introduces a spatial asymmetry in the magnitude of the relative phase difference between $\phi_{\perp}$ and $\phi_{\parallel}$ ($|\phi_{\perp}\!-\!\phi_{\parallel}|$). In the absence of phase difference, $|\phi_{\perp}\!-\!\phi_{\parallel}|$ is constant at $\pi/2$, resulting in the averaged intensity for one oscillation period being perfectly symmetric. However, when $\phi_{\mathrm{add}}$ is a finite value, such as $\pi/2$, $\phi_{\perp}$ and $\phi_{\parallel}$ are fixed in-phase on the $x\!<\!0$ and fixed out-of-phase on the $x\!>\!0$ region. Therefore, the intensity distribution is asymmetrical, reflecting the constructive and destructive interferences at the left- and right-hand sides.

Fig.~\ref{Fig:model} (c) illustrates the calculated spatial distributions of $|E^{CP}(x,y)|$ in typical $\phi_{\mathrm{add}}$ values. 
Considering that an identical break occurred in the symmetry of the electric field strength distribution on all four edges, the electric field distributions $|E^{CP} (0\!<\!\phi_{\mathrm{add}}\!<\!\pi)|$ exhibit a clear rotational displacement. 
The degree of asymmetry of the electric field increases until $\phi_{\mathrm{add}}$ reaches $\pi/2$, and then decreases to become isotropic at $\pi$. 
These calculations demonstrate that the essential factor for the rotationally displaced optical intensity is the phase difference between $E_{\perp}$ and $E_{\parallel}$.

This interpretation can hold true for complex intensity and phase distributions as well. We calculate the average intensity in the cases where $\phi_{\mathrm{add}}$ exhibits a curved phase distribution as observed in the FDTD simulation (Fig.~\ref{Fig:FDTD} (e)), and in the case where both phase and amplitude comprised complex distributions, assuming a superposition of higher-order bright modes. Although the intensity distributions are different, clear rotationally displaced patterns appear in both cases (Fig.~\ref{Fig:model} (d)).

\begin{figure}[t]
  \begin{center}
  \includegraphics[width=8.6cm]{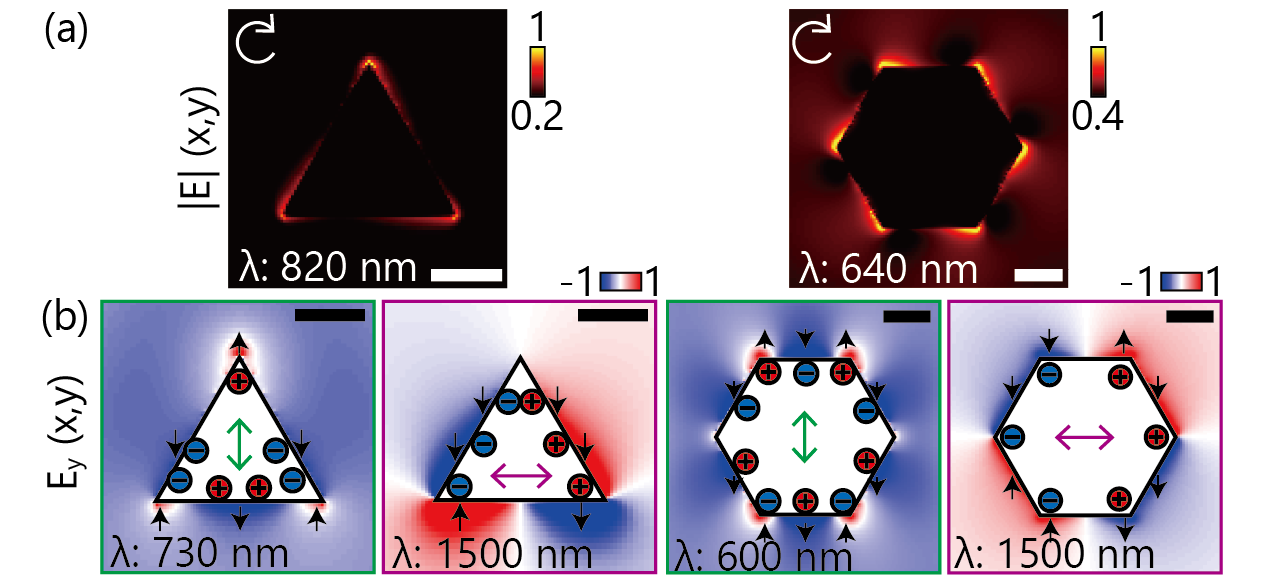}
  \end{center}
  \caption{(a) Spatial distributions of the electric field intensity for polygon plates excited by the circularly polarized (CP) light. (b) $E_{y}$ components at a representative phase at typical wavelengths. The edge lengths of the triangle and hexagon are 500 nm and 400 nm, respectively. Scale bars represent 200 nm.}
  \label{Fig:general}
\end{figure}

In the case of the square structure, the dipole and hexapole modes introduce phase differences for symmetry breaking. Although the specific modes and phase distributions depend on individual cases, the proposed interpretation of the origin of the rotationally displaced distribution should be valid for other regular polygons. Considering the superposition of distinct resonance modes induced by both perpendicular and horizontal polarizations with respect to one of the edges, asymmetric interference along the edges is expected~\cite{Horrer20NL}. As indicated in Fig.~\ref{Fig:general}, the FDTD simulations for triangular and hexagonal plates corroborate the distinct oscillation modes for the $y$- and $x$-polarized LP light, manifesting the rotational displacement under CP light at a wavelength between these modes.

Although the detailed calculations in this paper are concentrated on two-dimensional regular polygonal plates, extending this interpretation to a rectangular structure should be feasible by considering the phase differences between the major and minor axes, which define the vertices at which the electric field becomes localized. Constructing a three-dimensional model that considers the thickness of plates and the presence of a substrate, along with the experimental observation of rotationally displaced optical intensity, are among our future tasks. 

\section{Conclusion}
We numerically investigate the rotationally displaced intensity distributions around a square nanoantenna excited by CP light, using FDTD simulations and two model calculations. When the rotationally displaced pattern appeared, the electric fields at the square edges induced by the linear polarization components exhibit a concave phase distribution. This non-uniform phase distribution is explained by the superposition of dipole and hexapole modes, indicating that the rotationally displaced patterns require the involvement of multiple modes. The phase difference between edges perpendicular and parallel to the polarization direction caused by the non-uniform phase distribution is essential for the rotationally displaced intensity distributions around regular polygon antennas.

The proposed interpretation provides insights for discussing the chiral electromagnetic field derived from plasmonic nanostructures. Particularly, the non-uniform phase distribution generated by the superposition of multiple oscillation modes should be considered in calculating the optical chirality enhanced by plasmonic antennas. The potent relationship between the spatial phase distribution of a resonator excited by LP light and its symmetry is indispensable for understanding the interaction between the enhanced electric field of resonators and light–matter interactions.

\section*{Acknowledgments}
This work was supported by the JSPS KAKENHI (Grants No. JP23KJ0355, JP20K20560, No. JP20H00325)

\bibliography{diffraction}

\end{document}